\documentstyle[12pt,aasms4]{article}

\def\ltsima{$\; \buildrel < \over \sim \;$}
\def\simlt{\lower.5ex\hbox{\ltsima}}
\def\gtsima{$\; \buildrel > \over \sim \;$}
\def\simgt{\lower.5ex\hbox{\gtsima}}
\def\kms{\ifmmode {\rm \ km \ s^{-1}}\else $\rm km \ s^{-1}$\fi}
\def\mpc{$\ {h^{-1}\rm Mpc}$}
\def\bro{{{\bf r}_{{\scriptscriptstyle 1}}}}
\def\brt{{{\bf r}_{{\scriptscriptstyle 2}}}}
\def\mun{{{\mu}_{{\scriptscriptstyle N}}}}
\def\mutw{{{\mu}_{{\scriptscriptstyle 2}}}}
\def\mut{{{\mu}_{{\scriptscriptstyle 3}}}}
\def\mn{{m_{{\scriptscriptstyle N}}}}
\def\mo{{m_{{\scriptscriptstyle 1}}}}

\def\sd{\sigma_{{\scriptscriptstyle \Delta}}}
\def\sv{{\sigma_{{\scriptscriptstyle 21}}}}
\def\vs{{v_{{\scriptscriptstyle 21}}}}
\def\vl{{\bf v}^{{\scriptscriptstyle (1)}}}
\def\vq{{\bf v}^{{\scriptscriptstyle (2)}}}

\def\dl{{\delta^{{\scriptscriptstyle (1) }}}}
\def\dq{{\delta^{{\scriptscriptstyle (2) }}}}
\def\dlp{{\delta^{{\scriptscriptstyle (1) \,\prime}}}}

\def\zl{{v_z^{{\scriptscriptstyle (1)}}}}
\def\zlp{{v_z^{{\scriptscriptstyle (1) \,\prime}}}}

\def\zqp{{v_z^{{\scriptscriptstyle (2)\,\prime}}}}

\begin{document}

\title
{Skewed exponential pairwise velocities \\
from Gaussian initial conditions}
\author{Roman Juszkiewicz,\altaffilmark{1,2,3}
Karl B. Fisher\altaffilmark{3}
and Istv\'an Szapudi\altaffilmark{4,5}}

\altaffiltext {1} {Copernicus Astronomical Center, Bartycka 18,
00-716 Warsaw, Poland}
\altaffiltext {2} {University of Pennsylvania,
Department of Physics, Philadelphia, PA 19104-6396}
\altaffiltext {3}{Institute for Advanced Study,
School of Natural Sciences, Princeton, NJ 08540}
\altaffiltext {4} {NASA/Fermilab Astrophysics Center,
Batavia, IL 60510-0500}
\altaffiltext {5} {University of Durham, Physics Department, South Road,
Durham, DH1 3LE, United Kingdom (present address)}

\begin{abstract}

Using an Eulerian perturbative calculation, we show that the
distribution of relative pairwise velocities
which arises from gravitational instability of
Gaussian density fluctuations
has asymmetric (skewed) exponential tails.
The negative skewness is induced by
the negative mean streaming velocity of pairs (the infall
prevails over expansion), while
the exponential tails arise because
the relative pairwise velocity is a {\it number}, not
volume weighted statistic.
The derived probability distribution is compared
with N-body simulations and shown to provide a
reasonable fit.

\end{abstract}


\section{Introduction}
\label{sec-intro}

Redshift surveys present a distorted picture of the world
because peculiar motions displace galaxies from their true
spatial positions.
This phenomenon, which would make redshift surveys useless
for intergalactic spaceship navigators, is extremely useful
for cosmologists. It can
serve as a probe of the dynamics of gravitational
clustering and the cosmological mass density parameter, $\Omega$
(\cite{sar77}; \cite{pee80}, hereafter LSS;
\cite{kai87}; \cite{ham92}; \cite{pee93}, hereafter PPC;
\cite{reg95}).
A convenient statistical measure of the distortion effect is the
galaxy two-point correlation function in redshift space.
Under certain assumptions it can be expressed as
a convolution of the true spatial correlation function,
$\xi(r)$, with the distribution of the relative line-of-sight
velocities of pairs of galaxies, $p(w|r,\theta)$ .
Here $r$ and $w$ are respectively, the spatial separation and relative radial
velocity of a pair of galaxies, while $\theta$ is
the angle between the separation vector ${\bf r}$ and
observer's line of sight (cf. LSS; \cite{fis95},
hereafter F95). The purpose of this {\it Letter}
is to derive $p(w|r,\theta)$, using
weakly nonlinear gravitational instability theory.
This distribution was measured from N-body simulations and
estimated indirectly from redshift surveys. At $r \simlt$  1\mpc,
where\footnote{We use the standard parametrisation for the Hubble constant,
$H = 100 \, h$ \kms Mpc$^{-1}$.}
the galaxies are strongly clustered ($\xi \simgt 20$), the observations
are consistent with an exponential distribution (\cite{pee76};
\cite{dav83}; \cite{fis94}, hereafter F94; \cite{mar95};
\cite{lsd97}).
The fact that $p(w)$ at small separations
differs strongly from its initial, Gaussian character, is
not surprising: after all, the small-scale velocity field
has been `processed' by strongly-nonlinear dynamics in clusters,
and exponential distributions were recently derived from the
\cite{pre74} theory (\cite{she96}, \cite{dia96}).
On larger scales, where the fluctuations have
small amplitudes, one na{\"\i}vely expects to see the `unprocessed'
initial conditions. However, N-body experiments suggest that
$p(w|r,\theta)$ retains its exponential character even
at separations $r \simgt$ 10\mpc, where $\xi \simlt 0.1$,
despite the fact that the initial density and velocity fields
in those experiments were drawn from a Gaussian distribution
(\cite{efs88}, hereafter EFWD;
\cite{zur94}, hereafter ZQSW; F94).
At similar separations, an exponential $p(w|r,\theta)$ has also
been inferred from observations (F94; \cite{lov96}).
The simulations also
show that the radial component of the distribution,
$p(w|r,0^{\circ})$ is significantly skewed, in particular at large
separations (EFWD; ZQSW; F94).
The physical origin of the skewness and exponential shape of
$p(w)$ at large separations has until now remained unexplained.
We provide the explanation below.

\section{The origin of the negative skewness}
\label{sec-moments}

Let ${\bf v}$ and $\delta$ be the peculiar
velocity of a galaxy and the mass density contrast
at comoving position $\bro$,
while ${\bf v'}$ and $\delta'$ the velocity of another galaxy at position
$\brt$ at a certain fixed separation ${\bf r} = \brt -
\bro$. In our coordinate system
${\bf r} = \{x,y,z\} = r\{\sin\theta\cos\varphi, \,\sin\theta\sin\varphi,
\, \cos\theta\}$; the unit vector
$\, {\bf\hat z}$ points along the observer's line of sight, while
$v_z$ and $v'_z$ stand for the line of sight velocity components.
The probability that the four considered random fields
reach values
$\Re = \{\delta, \delta', {\bf v, v'} \}$ in the range
$d\Re = d\delta\, d\delta' \, d{\bf v} \, d{\bf v'}$ is
$g(\Re) \, d\Re$, and we will use brackets to denote ensemble averaging,
$\langle \, \ldots \, \rangle  = \int \ldots \,g\,d\Re$.
As usual, expectation values $\langle\ldots\rangle$ are
assumed to be equal to spatial averages. The latter, however, should
not be confused with number-weighted averages carried over galaxy
positions (cf. LSS and F95).
The $N$-th moment of the relative velocity,
weighted by the density of pairs of galaxies, is given by
\begin{equation}
\mn = {\langle (v'_z - v_z)^N(1 + \delta_g)(1 + \delta_g')\rangle
\over \langle(1 + \delta_g)(1 + \delta'_g)\rangle}
\; ,
\label{eqn-mN}
\end{equation}
where $\delta_g = (\delta \rho /\rho)_g$ is the contrast in
the number density of galaxies, which may differ from the
spatial fluctuations in the mass distribution. Here, we
ignore this potential difficulty and implicitly assume
$\delta_g({\bf r}) = \delta({\bf r})$.
A central moment of order  $N$ is given by
\begin{equation}
\mun = \langle\,(v_z' - v_z - \mo)^N (1 + \delta)
\, (1 + \delta') \, \rangle \, \left[ 1 + \xi(r)\right]^{-1}
\label{eqn-muN}
\end{equation}
To calculate the first few moments,
we shall expand the random fields in perturbative series,
$
{\bf v} = \vl + \vq +\ldots
$,
$
\delta = \dl + \dq +\ldots
$,
where each superscript describes the perturbative order
($\dq = O[\dl]^2$, etc.). We assume
that all linear order terms are described by a joint-normal
probability distribution. To lowest non-vanishing order,
$
\mo (r,\theta) =
\vs(r)\,\cos\theta + O(\xi^2)
$;
$
\mutw (r,\theta) =
\sv^2(r,\theta)
+ O(\xi^2)
$;
$
\mut (r,\theta) =
6 \, \langle \,\zqp\zl(\zl - 2\zlp)
\,\rangle \; + \; O(\xi^3)
$,
where $\xi(r) =
\langle\,\dl\,\dlp\,\rangle$ is the linear
correlation function, while
the mean streaming
velocity, $\vs$, and the pairwise velocity dispersion,
$\sv^2$, are given by
\begin{equation}
\vs(r)	=  - 2Hf(\Omega) \,
\int_0^r \, \xi(s) \, (s/r)^2 \, ds  \; ,
\label{eqn-vs}
\end{equation}
\begin{equation}
\sv^2 (r,\theta)
= 2\, \left[ \Pi(0) - \Pi(r) \cos^2 \theta
- \Sigma (r) \sin^2 \theta \right] \;,
\label{eqn-mu2}
\end{equation}
where $f(\Omega) \approx \Omega^{0.6}$, while
$\Pi$ and $\Sigma$ are the radial and transverse
components of the velocity correlation tensor,
related to $\xi(r)$ by equations (21.72)-(21.75) in
PPC (see also \cite{gor88} and \cite{gro89}).
Note that the leading terms in the expansions for
the first two moments come from linear perturbation theory.
The third moment is different.
In the early universe, linear perturbation
theory is sufficient, and to first order,
$\mut = 0$, in agreement with
the assumed Gaussian initial conditions, symmetric
about $\delta = 0$. However, gravitational
instability breaks the initial symmetry
(LSS; \cite{jus93}, \cite{ber95}, hereafter BJDB).
To calculate $\mut$, we need the second-order term for the
velocity field. It can be obtained by inverting the expression
for $\nabla\cdot\vq$, derived by BJDB.
For a curl-free flow, the resulting skewness is
\begin{equation}
\mut(r, \theta) = \sv^2(S_V + S_A)\,\cos\theta
\;,
\label{eqn-skew2}
\end{equation}
where the first term,
\begin{equation}
S_V = 3\vs(r)\,(\cos^2\theta + C\,\sin^2\theta)
; \;\; {\textstyle C(\Omega) \approx {3\over 7} \, \Omega^{-1/21}} ,
\label{eqn-sv}
\end{equation}
is induced by the mean streaming, while the second,
\begin{equation}
S_A =  6\left[\Sigma(r) - \Pi(r)  \right]
\,(1 + \cos^2\theta) \,(1 - C)/(Hfr) ,
\label{eqn-sa}
\end{equation}
comes from the anisotropy of the relative
velocity dispersion tensor. When ${\bf r}\perp{\bf\hat z}$
(both galaxies are in the plane of the sky), as well as
for zero separation, the skewness vanishes:
$\mut(r,90^{\circ}) = \mut(0,\theta) = 0$,
in agreement with symmetry requirements.
Such requirements do not apply to the radial
component of the distribution (${\bf r} ~\|~ {\bf\hat z}$)
when the mean infall velocity is different from zero.
We will now show that if
$\xi$ remains positive for separations $< r$,
then $\mut(r,0^{\circ}) < 0$. First, note that
a positive $\xi$ implies $\vs(r) < 0$ and
$\Sigma(r) > \Pi(r)$ (see eq.~[\ref{eqn-vs}] above
and eqs. [21.72]-[21.75] in PPC). Hence, $S_V(r,0^{\circ}) < 0$
while $S_A(r,0^{\circ}) > 0$: the effect
of the negative infall velocity is counterbalanced
by the anisotropy of the velocity correlation tensor
(in agreement with N-body results; see p. 938 in F94).
However, the two effects do not cancel out and for
all `reasonable' models the term, induced by the
mean streaming dominates. For a power-law correlation function,
$\xi(r) \propto r^{-\gamma}$, the ratio of the two
terms is $S_A(r,0^{\circ})/S_V(r,0^{\circ}) = -8/7(5 - \gamma)
\approx - 0.35$ for $\gamma = 1.7$.

In order to test our perturbative predictions against fully nonlinear
N-body experiments, we used data, generated from the simulations
described in Frenk et al. (1990; appropriate codes were kindly
provided to us by Marc Davis). The simulations are of a standard
CDM model with $\Omega = 1, \, h = 0.5$,
and $\sigma_8 = 0.6$ (here $\sigma_8$ is the rms density contrast
in a 8 \mpc~sphere). These simulations
contain N$= 64^3$ particles in a $L = 180$ \mpc~box. The first
three moments at separation $r = 10.4$ \mpc, and $\theta = 0^{\circ}$,
determined directly from the simulations, are
$(\vs, \, \sv, \, \mut^{1/3})\,=\,(-190,\, 440, \, - 400)$ \kms,
while
equations~(\ref{eqn-vs}), (\ref{eqn-mu2}) and~(\ref{eqn-skew2}) give
$(-200, 430, -430)$ \kms, respectively. Comparisons
with numerical experiments of higher resolution (N $= 256^3$,
\cite{spr98}) suggest that the accuracy of
the N $= 64^3$ results, quoted above, is $\sim 20\%$. We conclude
that the perturbative predictions are in excellent agreement
with the N-body experiments.

\section{The origin of the exponential tails}
\label{sec-tails}

Eulerian perturbation theory, truncated at second order,
can be used to write $p(w|r,\theta)$ as a marginal probability,
obtained by integrating a 14-dimensional Gaussian distribution.
Here we will not do that, however. Instead, we will
trade accuracy for simplicity
of calculations and replace rigorous perturbation theory with
a following Ansatz. Suppose that the relation between the pairwise velocity
and the random vector $\Re = \{\delta, \delta', {\bf v, v'} \}$
is given by the mapping
\begin{equation}
w  =  u(1+\dlp)(1+\dl) \; =
\; u + u\Delta + O(u^3)
\; ,
\label{eqn-wdefn}
\end{equation}
where $u \equiv \zlp - \zl$ and
$\Delta \equiv \dlp + \dl$.
The first three moments of
this new variable can be readily
expressed in terms of $\vs$ and $\sv$. To lowest non-vanishing
order, we obtain
$
\langle w\rangle  =  \vs (r) \,\cos\theta$;
$\;
\langle (w - \langle w\rangle)^2\rangle =  \sv^2(r,\theta)
$, and
\begin{equation}
\langle (w - \langle w\rangle)^3\rangle  \; =  \;
6\,\sv^2 (r, \theta) \, \vs \, \cos\theta
\; .
\label{eqn-ansatz}
\end{equation}
Clearly, the transverse ($\theta = 90^{\circ}$) components of the above
moments agree with those obtained from rigorous second-order
Eulerian theory in \S~\ref{sec-moments} above.
Hence, our Ansatz provides an acceptable approximation
of the second-order prediction for $p(w|r,90^{\circ})$.
For the radial part, we recover the true values of the
first two moments only.
The third moment, obtained from the approximate distribution is
an overestimate of the true $|\mut(r,0^{\circ})|$.
However, the approximate expression ~(\ref{eqn-ansatz})
correctly reproduces the negative sign of $\mut$, the scaling with
the cosmological density parameter, $\mut \propto \Omega^{1.8}$,
as well as the scaling with the two lower moments, introduced by the
dominant, infall term $\propto \vs\sv^2$.

According to equation~(\ref{eqn-wdefn}),
$w$ is the sum of two variables, $u$ and $\varpi\equiv u\Delta$.
The probability distribution for $u$ is
\begin{equation}
p_u(u) = {{1}\over{\sqrt{2\pi}\sv}}\,
\exp\left\{ - {{u^2}\over{2\sv^2}}\right\} \; .
\label{eq-pu}
\end{equation}
The probability distribution for $\varpi \equiv u\Delta$
is readily obtained by integrating the expression
\begin{equation}
p_{\varpi}(\varpi|r,\theta) = \int\limits_{-\infty}^{+\infty}\,
{{du}\over{|u|}}\, p_{u\Delta}\left(u,{{\varpi}\over{u}}
\right) \; ,
\label{eqn-uDl}
\end{equation}
where $p_{u\Delta}(u,\varpi/u)$ is a standard, joint-normal
distribution for $u$ and $\Delta$ (eg. F95), with $\varpi/u$
substituted for $\Delta$. This integral gives
\begin{equation}
p_{\varpi}(\varpi|r,\theta)
= {{\alpha}\over{\pi\sv}}\,
e^{\beta\kappa\varpi}\,
{\rm K}_0\left( \beta |\varpi|\right) \; ,
\label{eq-probx}
\end{equation}
where ${\rm K}_0$ is the usual modified Bessel function,
$\alpha \equiv 1/\sd\sqrt{1-\kappa^2}$,
$\beta \equiv 1/\sv\sd(1-\kappa^2)$,
$\kappa \equiv \vs\cos\theta/\sd\sv$, and
$\sd^2 \equiv \langle\Delta^2\rangle =$
$2\left[\,\xi(0) + \xi(r)\,\right]$.
There are several remarks worth noting about the properties
of the random variable $\varpi$. Near the origin, its distribution has
a cusp; for small values
of the argument, ${\rm K}_0 (|x|) = - [\ln(|x|/2) + 0.577]
+ O(x^2)$. For large values of the argument,
${\rm K}_0(|x|) = \sqrt{\pi/2|x|}\,
e^{-|x|} \, [1 + O(1/|x|)]$: this is the exponential behavior
in the wings, typical for products of
Gaussian fields (eg. the $\chi^2$
distribution; see also \cite{sch92} or \cite{hol93}).
Finally, note that the exponential in equation~(\ref{eq-probx})
is not symmetric about $\varpi=0$, and its skewness
is introduced by cross-correlation between the
velocity and density (i.e., by the infall).
The distribution
of $w = \varpi + u$ is qualitatively similar to $p_{\varpi}$;
it can be obtained from an expression, similar
to eq.~(\ref{eqn-uDl}) with the original
integrand, replaced by $|u|^{-1}p_{u\Delta}\left[
u, (w/u) - 1 \right]$. The resulting integral can be
rewritten as
\begin{equation}
p(w|r,\theta) = \int\limits_{0}^{\infty}\,
{d\sigma\over\sigma}\,\exp\left\{-{w^2\over{2\sigma^2}}\right\}\,
W(w, \vs, \sv)\, ,
\label{eqn-probw}
\end{equation}
with
\begin{equation}
W = {{\alpha}\over \pi\sv}\,e^{\,\beta\kappa w
- {{1}\over{2}}\,(\beta^2\sigma^2 +\alpha^2)}\,
{\rm cosh}\left[\alpha\left({w\over{\sigma}} -
\kappa \beta\sigma\right)\right].
\label{eqn-ww}
\end{equation}
One can approximate the above integral via the method
of steepest descent.  This yields a convenient analytic
formula for the probability distribution,
\begin{equation}
p(w|r,\theta)\approx
{{\rm cosh}\left(
\sqrt{|U|}\,/K\sd\right)
\over{\sqrt{2\pi\sv\sd|w|}}} \,
\exp\left\{-{|U|\over K}-{\alpha^2\over 2}\right\}\,
\label{eq-sd}
\end{equation}
where $U = w/\sv\sd$, $K = 1+ {\rm sgn}(w)\kappa$,
${\rm sgn}(w)= +1$ for $w\ge 0$, and
${\rm sgn}(w)= -1$ for $w< 0$.
The characteristic function is given by the Fourier transform
of equation~(\ref{eqn-probw}),
\begin{equation}
\langle e^{iwt}\rangle \; = \;
[\phi(t)]^{-1/2}\, \exp\left( - \sv^2 t^2/2\phi(t)\right)
\; ,
\label{eq-char}
\end{equation}
where
$
\phi(t) = 1 -2i\kappa\sv\sd t
+ \sd^2(1-\kappa^2)\sv^2t^2 \, .
$
In the limit
$\sd\to 0$, $\langle e^{iwt}\rangle
\to \exp(- \sv^2t^2/2)$,
and we simply recover the Gaussian
distribution $p_u(u)$ (eq.~\ref{eq-pu}), in agreement
with F95.
This is obvious when we recall that non-Gaussian behavior
of the distribution arises from the quadratic nonlinearities
introduced by the number weighting; these
quadratic terms become small compared to the volume weighted
(and Gaussian distributed)
velocity difference as the amplitude of the fluctuations decreases.
In the opposite limit, when $\sd$ is increased, the distribution
rapidly develops a central cusp and exponential wings.
It is interesting to compare our results, valid in the transition
zone between the linear and nonlinear regime, on separations
$r \simgt$ 5 \mpc, with the analysis of
Diaferio \& Geller (1996) and Sheth (1996), restricted to much
smaller scales, where $\sd \gg 1$ and the perturbative approach
breaks down. Note that the integral in equation~(\ref{eqn-probw})
is a weighted sum over Gaussians having
a range of dispersions. The weighting factor $W$ is determined
by the velocity correlation tensor and the velocity-density
cross correlation, $\vs$. This expression is similar to equation (5)
of Sheth (1996), valid in the strongly nonlinear regime,
where $W$ is related to the Press-Schechter multiplicity function.
The outcome of summing Gaussian distributions in both cases
is an exponential distribution. A significant difference
is that in the strongly nonlinear regime all sources of the
velocity skewness vanish: the limit $r \to 0, \, \sd \gg 1$
corresponds to virialized cluster centers; there is
no infall ($\vs = 0$); the velocity dispersion
is isotropic $(\Pi - \Sigma=0)$, and $\mut = 0$
by symmetry.

We will now compare the predicted velocity distribution with direct
measurements from simulations. Since the N-body results we have
at our disposal assume a CDM spectrum, we need to introduce a
shortwave cutoff in the initial power spectrum; otherwise
$\xi(0)$ becomes infinite. We use a
Gaussian filter of width $R_s$ and multiply the linear
power spectrum, $\,P(k) = \int\xi(r)\exp(i{\bf k \cdot r})d^3r\;$
by $\;\exp(-k^2R_s^2)$. The resulting $\xi(r)$ is finite at $r =
0$ and remains flat for $r \simlt R_s$. Existence of such
a `core radius' in any realistic $\xi(r)$ is necessary
anyway since galaxies are not point-like objects and have
finite sizes. $R_s$ can also be related to the effective
dynamical resolution of the simulations; we postpone the
discussion of this problem to a later paper
(Springel et al. 1998). Finally, the small-scale cutoff
can be useful as a makeshift solution for
reducing the $|\mut|$, overestimated by our
Ansatz, and bringing this moment closer to the value predicted
by the rigorous perturbative calculation in \S 2 above.
We tried several filtering widths and found that $R_s =$
3 \mpc~provides a reasonable fit to N-body simulations.
In Figure 1 we compare the
probability distribution, $p(w|r,0^{\circ})dw$, calculated
from equation~(\ref{eqn-probw}), with direct measurements
from N-body simulations. The upper panel shows the results of
measurements from the simulations of Frenk et al. (1990;
N $\approx 2.6 \times 10^5$),
while the lower panel -- those from {\.Z}urek et al. (1994;
N $\approx 1.7 \times 10^7$), obtained by fitting their Fig. (7d)
by a double exponential. The separation is respectively
$r = 10.4$ \mpc, and $r = 10.5$ \mpc.
The sizes of the velocity bins
are $dw =$ 72 and 100 km s$^{-1}$ for the upper and lower panel,
respectively. The
assumed $P(k)$ in both cases is the standard CDM
spectrum, described in \S2 (the only difference is that here
we use $P(k)\exp(-k^2R_s^2)$ with $R_s = 3$ \mpc,
while in \S 2 $R_s = 0$).
Clearly, the agreement between the perturbative predictions
and N-body measurements improves when the resolution
of the simulations is improved.

A possible alternative to the Ansatz, adopted in
this section, is to derive $p(w)$ from the
\cite{zel70} approximation (hereafter ZA).
Like our Ansatz, this approach makes
calculations simpler than the rigorous treatment
in \S 2 above, and the simplicity here, too,
is bought at the expense of accuracy in estimating
$\mut$: at second
order, the ZA breaks the momentum conservation
(\cite{jus93}; BJDB) by implying $C = 0$ in eqs.~(\ref{eqn-sv})
and~(\ref{eqn-sa}).
As a result, the ZA underestimates $|\mut|$ by $\sim$ 50\%.
At first order, however, the ZA agrees with the rigorous
Eulerian perturbation theory, so its predictions
for $\vs$ and $\sv$ must be identical with ours. The ZA
was recently used by \cite{set97}.
Qualitatively, their $p(w)$ is similar to ours.
However, at the quantitative level we disagree
because their method underestimates $\vs$
by at least an order of magnitude. As a consequence,
Seto and Yokoyama had to readjust their predicted
$p(w)$ `by hand' to achieve agreement with simulations.
Their results seem puzzling given the
properties of the ZA, discussed above.

Another alternative approach one might consider,
is to expand $p(w)$ in orthogonal polynomials
(eg. \cite{jus95}; \cite{lif81}, p. 31). At the end,
only direct applications to future redshift
surveys, like the SDSS or 2dF, will decide which
of the discussed physical models of $p(w)$ provides
the optimal combination of simplicity and accuracy.

We thank  M. Chodorowski, F. Bouchet, J. Frieman,
M. Ruszkowski, R. Scherrer, R. Stompor, V. Springel and S. White
for stimulating discussions. This work began
in the creative atmosphere of the Aspen Center for Physics
and we thank the Organizers of the Summer meetings
there in 1994 and 1997. Our research was supported by
grants from the Polish Government (KBN grants 
No. 2.P03D.008.13 and 2.P03D.004.13),
the Ambrose Monell Foundation, the DoE,
NASA (NAG-5-2788) and by the Poland-US
M. Sk{\l}odowska-Curie Fund.

\clearpage

\noindent{\bf FIGURE CAPTION}

\noindent{\bf Figure 1}

Comparison of the probability distribution,
$dP = p(w|r,0^{\circ})dw$, derived from equation~(\ref{eqn-probw})
(shown as curves) with direct determinations from N-body experiments
(shown as histograms). The number of particles in the two sets
of simulations was N = $2.6 \times 10^5$ ({\it upper panel}) and
N = $1.7 \times 10^7$ ({\it lower panel}). Note that an increase
in resolution brings the N-body results closer to the
perturbative predictions.

\begin{thebibliography}{}
\bibitem [Bernardeau et al. 1995] {ber95}
Bernardeau, F., Juszkiewicz, R., Dekel, A.
\& Bouchet, F.R. 1995, \mnras, 274, 20 (BJDB)
\bibitem [Davis and Peebles 1983] {dav83}
Davis, M. \& Peebles, P.J.E 1983, \apj, 267, 465
\bibitem [Diaferio and Geller 1996]
{dia96} Diaferio, A. \& Geller, J.M. 1996, \apj, 467, 19
\bibitem [Efstathiou et al. 1988] {efs88}
Efstathiou, G., Frenk, C., White, S., \&
Davis, M. 1988, \mnras, 235, 715 (EFWD)
\bibitem [Fisher 1995] {fis95}
Fisher, K.B. 1995, \apj, 448, 494 (F95)
\bibitem [Fisher at al. 1994] {fis94}
Fisher, K.B, et al. 1994, \mnras, 267, 927 (F94)
\bibitem [Frenk et al. 1990]{fre90}
Frenk, C., White, S., Efstathiou, G.,
\&  Davis, M. 1990, \apj, 351, 10
\bibitem [G{\'o}rski 1988]{gor88}
G{\'o}rski, K. 1988, \apjl, 332, L7
\bibitem [Groth et al. 1989]{gro89}
Groth, E.J., Juszkiewicz, R., \& Ostriker, J.P. 1989,
\apj, 346, 558
\bibitem [Hamilton 1992] {ham92}
Hamilton, A. J. S. 1992, \apj, 385, L5
\bibitem [Holzer and Siggia 1993]{hol93}
Holzer, M., \& Siggia, E. 1993, Phys. Fluids A, 5 (10),
2525
\bibitem [Juszkiewicz et al. 1993]{jus93}
Juszkiewicz, R., Bouchet, F.R., \& Colombi, S. 1993,
\apjl, 412, L9
\bibitem [Juszkiewicz et al. 1995]{jus95}
Juszkiewicz, R., et al. 1995, \apj, 442, 39
\bibitem [Kaiser 1987] {kai87}
Kaiser, N. 1987, \mnras, 227, 1
\bibitem[Landy, Szalay, \& Broadhurst 1997]{lsd97}
Landy, S.D., Szalay, A.S., Broadhurst, T.J. 1997,
\apjl, submitted (astro-ph/9711045)
\bibitem [Lifshitz and Pitaevskii 1981] {lif81}
Lifshitz, E.M., \& Pitaevskii, L.P. 1981,
Physical Kinetics, (Oxford: Pergamon Press)
\bibitem [Loveday et al. 1996] {lov96}
Loveday, J., Efstathiou, G., Maddox, S.J.,
\& Peterson, B. A. 1996, \apj, 468, 1
\bibitem [Marzke et al. 1995] {mar95}
Marzke, R.O., Geller, M.J., da Costa, L.N., \& Huchra,
J.P. 1995, \aj, 110, 477
\bibitem [Peebles 1976] {pee76}
Peebles, P.J.E. 1976, \apss, 45, 3
\bibitem [Peebles 1980] {pee80}
Peebles, P.J.E. 1980, The Large-Scale Structure of the Universe,
(Princeton:  Princeton University Press) (LSS)
\bibitem [Peebles 1993] {pee93}
Peebles, P.J.E. 1993, Principles of Physical Cosmology,
(Princeton:  Princeton University Press) (PPC)
\bibitem [Press-Schechter (1974)] {pre74}
Press, W., \& Schechter, P. 1974, \apj, 187, 425
\bibitem [Reg{\"o}s and Szalay 1995] {reg95}
Reg{\"o}s, E., \& Szalay, A. 1995, \mnras, 272, 447
\bibitem [Sargent and Turner 1977] {sar77}
Sargent, W. W., \& Turner, E. L. 1977, \apjl, 212, L3
\bibitem [Scherrer 1992] {sch92}
Scherrer, R.J. 1992, \apj, 390, 330
\bibitem [Seto and Yokoyama (1998)] {set97}
Seto, N., \& Yokoyama, J. 1998, \apj, 492, 421 
\bibitem [Sheth 1996] {she96}
Sheth, R.K. 1996, \mnras, 279, 1310
\bibitem [Springel et al. 1998] {spr98}
Springel, V., Juszkiewicz, R., \& White, S. 1998,
paper in preparation
\bibitem [Zel'dovich (1970)] {zel70}
Zel'dovich, Ya.B. 1970, \aap, 5, 84
\bibitem [{\.Z}urek et al. 1994] {zur94}
\.Zurek, W.H., Quinn, P.J., Salmon, J.K., \& Warren, M.S.
1994, \apj, 431, 559 (ZQSW)
\end{thebibliography}
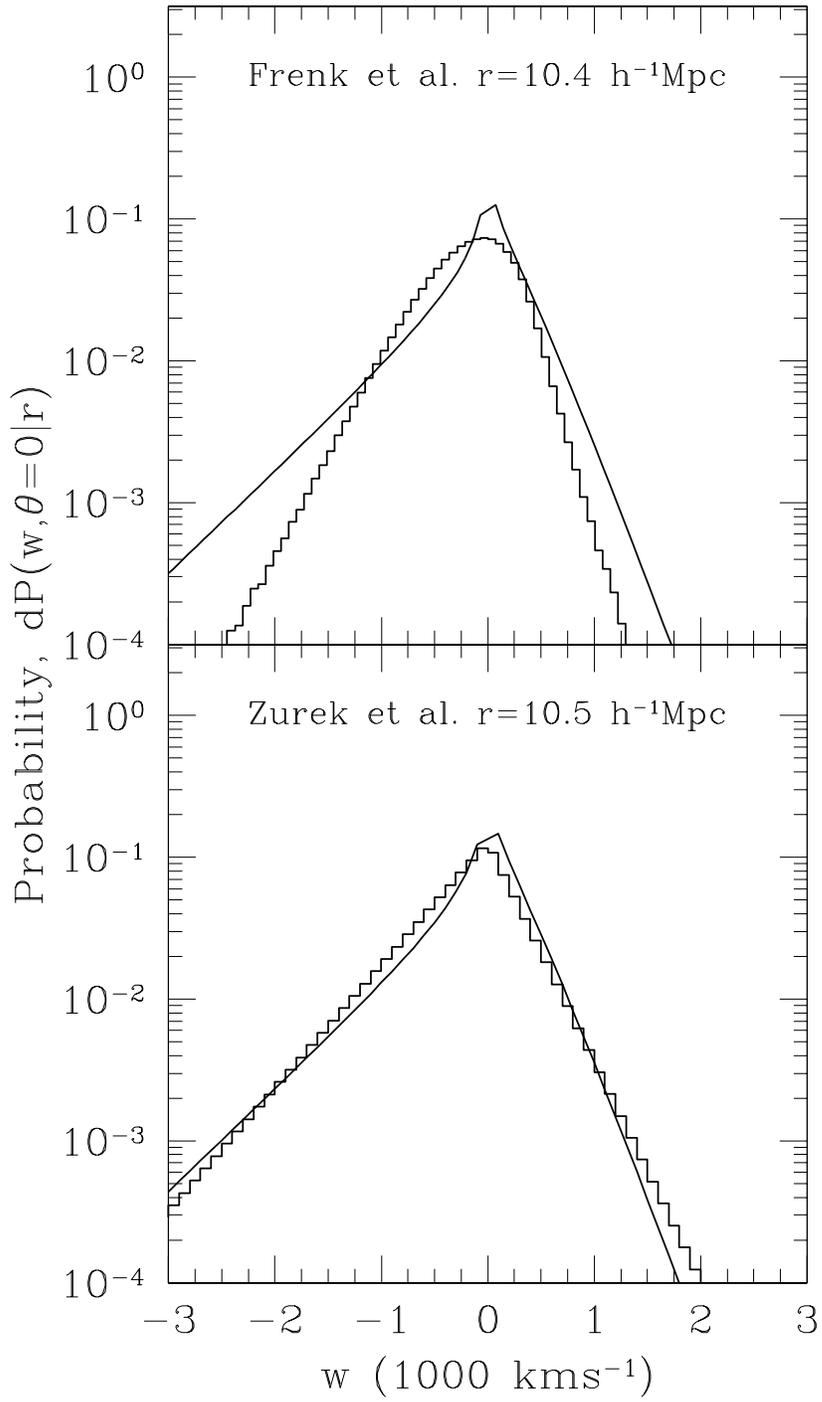
\end{document}